\documentclass[a4paper,12pt,preprint]{elsarticle}
\usepackage{amssymb}
\usepackage{amsfonts}
\usepackage{amsmath}

\usepackage{graphicx}
  \setkeys{Gin}{width=\linewidth}

\journal{arXiv.org}

\biboptions{sort&compress}

\begin{document}

\begin{frontmatter}

\title{Effects of transverse electric fields on Landau subbands in bilayer zigzag graphene nanoribbons}


\author[NCKU]{Hsien-Ching Chung}
\ead{hsienching.chung@gmail.com}
\author[NCHPC]{Po-Hua Yang}
\ead{yangp@narlabs.org.tw}
\author[KunShan]{To-Sing Li}
\ead{tsli@mail.ksu.edu.tw}
\author[NCKU]{Ming-Fa Lin}
\ead{mflin@mail.ncku.edu.tw}


\address[NCKU]{Department of Physics, National Cheng Kung University, Tainan 70101, Taiwan}
\address[NCHPC]{National Center for High-Performance Computing (South), Tainan 74147, Taiwan}
\address[KunShan]{Department of Electrical Engineering, Kun Shan University, Tainan 71003, Taiwan}

\begin{abstract}
The magnetoelectronic properties of quasi-one-dimensional zigzag graphene nanoribbons are investigated by using the Peierls tight-binding model. Quasi-Landau levels (QLLs), dispersionless Landau subbands within a certain region of $k$-space, are resulted from the competition between magnetic and quantum confinement effects. In bilayer system, the interlayer interactions lead to two groups of QLLs, one occurring at the Fermi level and the other one occurring at higher energies. Transverse electric fields are able to distort energy spectrum, tilt two groups of QLLs and cause semiconductor-metal transition. From the perspective of wave functions, the distribution of electrons is explored, and the evolution of Landau states under the influence of electric fields is clearly discussed. More interestingly, the band mixing phenomena exhibited in the energy spectrum are related to the state mixing, which can be apparently seen in the wave functions. The density of states, which could be verified through surface inspections and optical experiments, such as scanning tunneling spectroscopy and absorption spectroscopy, is provided at last.
\end{abstract}


\end{frontmatter}


\section{Introduction}

Graphene, a truly two-dimensional system with carbon atoms arranged in the honeycomb lattice, was first discovered in 2004~\cite{Science306(2004)666K.S.Novoselov}.
It has many unique properties, such as linear energy spectrum at the Dirac point~\cite{Phys.Rev.71(1947)622P.R.Wallace}, fractional quantum hall effect~\cite{Nature438(2005)197K.S.Novoselov, Nature438(2005)201Y.B.Zhang}, and Klein tunneling~\cite{Rev.Mod.Phys.80(2008)1337C.W.J.Beenakker, Phys.Rev.Lett.102(2009)026807N.Stander}.
The ambipolar and high carrier mobility properties stimulated many imaginations in electronic applications~\cite{Synth.Met.161(2011)2249Y.Y.Tan}.
Graphene nanoribbons, as an alternative, are obtained from cutting graphene along a specified direction.
There are two main types of nanoribbon, namely zigzag and armchair, according to the shape of the confining edges.
They present quite different electronic properties.
The zigzag ones belong to a gapless semiconductor, since the strongly localized edge states are lying at the Fermi level ($E_F=0$).
The armchair ones are semiconducting or gapless material depending on the ribbon width.
Practically, graphene nanoribbons are available from the chemical route~\cite{Science319(2008)1229X.L.Li} or the longitudinal unzipping of carbon nanotubes~\cite{Nature458(2009)872D.V.Kosynkin}, and their peculiarities have drawn many interesting studies on electronic and magnetic properties~\cite{Rev.Mod.Phys.81(2009)109A.H.CastroNeto, Phys.Rev.B59(1999)8271K.Wakabayashi, J.Phys.Soc.Jpn.80(2011)044602H.C.Chung, Nanotechnology18(2007)495401Y.C.Huang, PhysicaE42(2010)711H.C.Chung}, optical properties~\cite{Opt.Express19(2011)23350H.C.Chung, J.Phys.Soc.Jpn.69(2000)3529M.F.Lin}, electronic excitations~\cite{J.Phys.Soc.Jpn.70(2001)2513M.F.Lin}, and transport properties~\cite{Phys.Rev.B76(2007)205433A.Cresti, Synth.Met.171(2013)7T.S.Li}.

The low-energy energy spectrum of monolayer zigzag graphene nanoribbon consists of doubly degenerate partial flat bands and many parabolic energy subbands~\cite{J.Phys.Soc.Jpn.80(2011)044602H.C.Chung}.
The former, coming from the edge states, lie at the Fermi level within the region $2\pi/3 < k_x \leq \pi $.
The latter are resulted from the quantum confinement effects of the finite width of ribbon.
The conduction and valence parabolic subbands occurring from the Fermi level are symmetric to each other about $E_F=0$.
A perpendicular magnetic field would induce Landau subbands in graphene~\cite{Carbon42(2004)2975C.P.Chang} which brings the states with similar energies together.
Due to the finite width of ribbon, the Landau subbands, being dispersionless within a certain range of wavevector, are called ``quasi-Landau levels (QLLs).''

For bilayer graphene nanoribbons under the influence of magnetic fields, there are two groups of Landau subbands and two pairs of doubly degenerate partial flat bands resulting from interlayer interactions.
The first group of Landau subbands occurs from the Fermi energy, while the second group from higher energies.
The two pairs of flat subbands are lying on different energies.

In this work, a transverse electric field, which cause the electrons to experience a smoothly changed electric potential along the ribbon, is taken into consideration.
The electronic properties are drastically changed, such as distortion of Landau subbands, destruction of dispersionless feature, and splitting of the degenerate flat subbands.
In addition, band mixing phenomena appear and cause mixing of states.
These phenomena are investigated and realized by analyzing wave functions.
The density of states (DOS), closely related to the available channels in optical excitations and transport properties, could be measured through surface inspections and optical experiments, such as scanning tunneling spectroscopy and absorption spectroscopy.
They are discussed in the last section.

\section{Theory}

\begin{figure}
\begin{center}
  \includegraphics[]{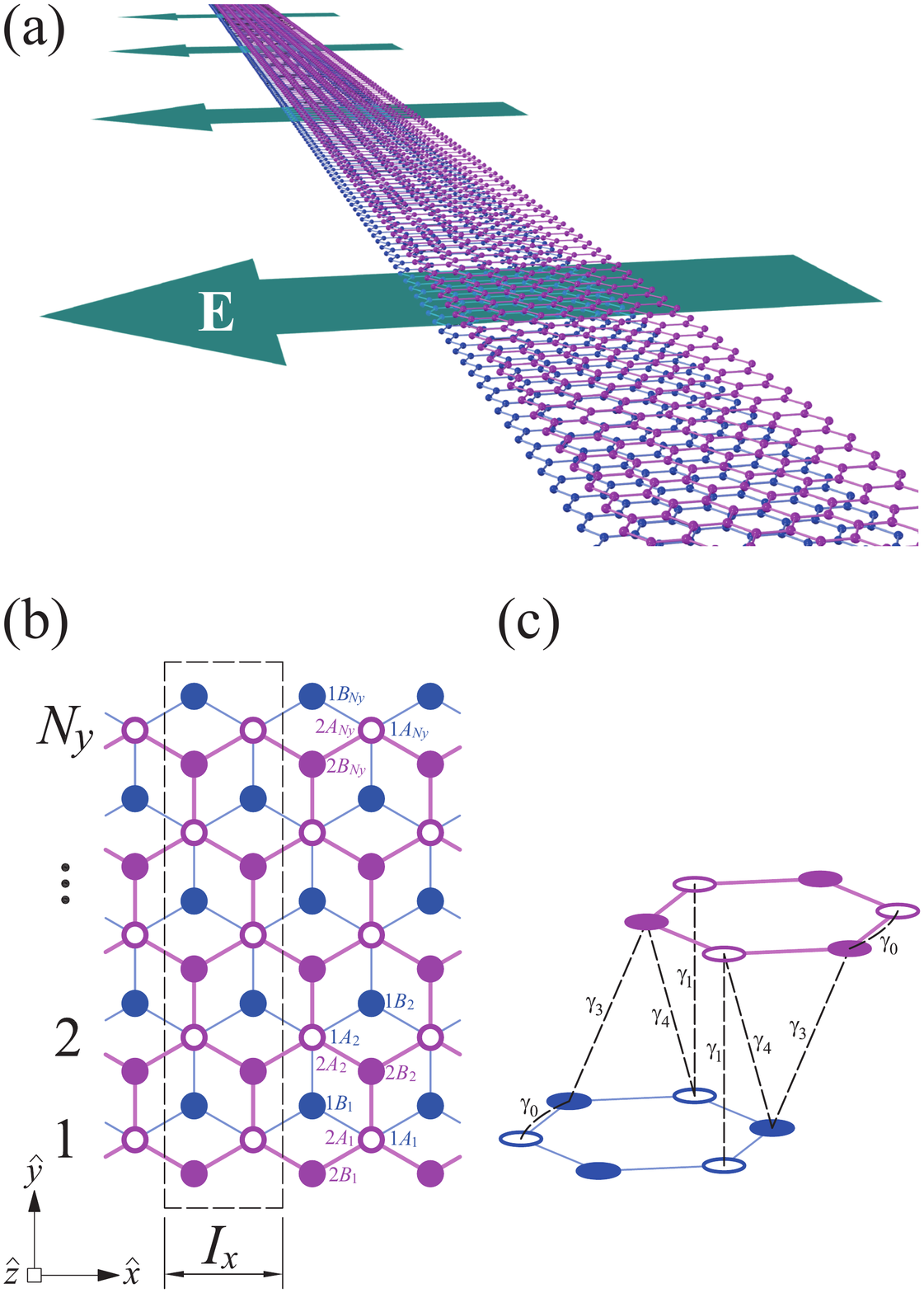}\\
  \caption{
  (a) An AB-stacked bilayer zigzag graphene nanoribbon in external fields. The two layers of graphene nanoribbon are rendered in different colors.
  (b) Geometric structure of bilayer zigzag graphene nanoribbon.
  (c) The hopping integrals $\gamma_0$, $\gamma_1$, $\gamma_3$, and $\gamma_4$ are indicated by the dashed-lines ($\gamma_6$ representing the chemical shift is not included in the figure).
  }
  \label{fig:EYBZ2011GeometricStructure}
\end{center}
\end{figure}

The illustration about an AB-stacked bilayer zigzag graphene nanoribbon in external fields is shown in Fig.~\ref{fig:EYBZ2011GeometricStructure}(a).
The magnetic field $\mathbf{B} = ( 0, 0, B_\perp )$ is along the $z$-direction (perpendicular to the ribbon plane) and the transverse electric field $\mathbf{E} = ( 0, E_\parallel , 0 )$ is along the $y$-direction.
A detailed top-view for the geometric structure is depicted in Fig.~\ref{fig:EYBZ2011GeometricStructure}(b).
The zigzag graphene nanoribbons, which are the strip of carbon atoms arranged in a honeycomb (hexagonal) lattice with zigzag edges, lie on the $xy$-plane.
Atoms $A$ and $B$, denoted by circles ($\circ$) and dots ($\bullet$) respectively, are the nearest neighbors to each other with a C--C bond length $b=1.42$ \AA , and the distance between ribbon layers is 3.35 \AA .
The width of nanoribbon is $W=(3N_{y}-1)b/2$, where $N_{y}$ is the number of zigzag lines along the $y$-direction.
The number in front of $A$ ($B$) denotes which layer the atom locates, and the subscript stands for the $m$-th zigzag line where the atom locates.
For example, $1A_2$ represents the $A$ atoms on the 2nd zigzag line of the 1st ribbon layer.
The dashed-line rectangle represents the primitive unit cell with a periodical length $I_x=\sqrt{3}b$, and the first Brillouin zone is in the region $|k_{x}| \leq \pi $.
There are $4N_{y}$ carbon atoms in a unit cell.

The tight-binding method is used to study the low-energy electronic properties in the presence of external electric and magnetic fields.
In this work, only the $\pi$ electrons are taken into consideration, and the hopping integrals (atom-atom interactions) are described as follows [Fig.~\ref{fig:EYBZ2011GeometricStructure}(c)]: $\gamma _0$ is the intralayer interaction between atoms $A$ and $B$. $\gamma_1$ ($\gamma_3$) represents the interlayer interaction between two $A$ ($B$) atoms.
$\gamma_4$ corresponds to the interlayer interaction between atoms $A$ and $B$. $\gamma_6$ is the chemical-shift between $A$ and $B$.
The values of $\gamma_i$'s in AB-stacked graphite are adopted in our calculation: $\gamma_0=2.598$ eV, $\gamma_1=0.364$ eV, $\gamma_3=0.319$ eV, $\gamma_4=0.177$ eV, and $\gamma_6=-0.026$ eV~\cite{Phys.Rev.B43(1991)4579J.C.Charlier}.

The wave function is the linear superposition of the $4N_y$ tight-binding functions.
\begin{equation}
|\Psi (k_{x})\rangle = \sum\limits_{m=1}^{N_{y}}
\bigg[ 1A_m(k_{x}) |a_m\rangle + 1B_m(k_{x}) |b_m\rangle
+ 2A_m(k_{x}) |a_m\rangle  + 2B_m(k_{x}) |b_m\rangle \bigg] ,
\label{eq:EYBZ2011BilayerWF}
\end{equation}
where $1A_m(k_x)$ [$1B_m(k_x)$] is site amplitude of atom $A$ ($B$) on the $m$-th line of the 1st layer at wavevector $k_x$ and $2A_m(k_x)$ [$2B_m(k_x)$] for the 2nd layer.
$|a_m\rangle $ ($|b_m\rangle $) is the Bloch tight-binding function of $2p_z$ orbitals associated with the periodic $A$ ($B$) atoms.

The transverse uniform electric field $\mathbf{E}=E_\parallel \hat{y}$ with a maximal field strength of $0.2$ mV/{\AA} is taken into account.
The geometric structure and the tight-binding parameters are little affected by the electric field with such a strength.
Thus, the effect of electric field can be treated as a perturbation term $U=-eE_\parallel y$ ($y$ is the $y$-component of atom's position) added to the on-site energy in the Hamiltonian.

When a perpendicular magnetic field $\mathbf{B}=B_\perp \hat{z}$ is applied along the $z$-direction, the Hamiltonian matrix elements can be given by multiplying the zero field matrix elements with a phase factor, $\exp (i 2\pi \theta_{ll'})$~\cite{PhysicalPropertiesOfCarbonNanotubes(1998)p98R.Saito}, where $\theta_{ll'}=(1/\phi _0)\int_{\mathbf{R}_l}^{\mathbf{R}_{l'}}\mathbf{A}(\mathbf{l})\cdot d\mathbf{l}$ is the so-called Peierls phase, a line integral of the vector potential $\mathbf{A}$ from the $l$-th to $l'$-th atom [$\mathbf{R}_{l}$ ($\mathbf{R}_{l'}$) is the position vector of $l$-th ($l'$-th) atom] in the unit of the magnetic flux quantum $\phi _0=h/e$.
The vector potential is chosen as $\mathbf{A}=(-B_\perp y,0,0)$ in Landau gauge, which preserves the translation invariance along the $x$-axis.

The Hamiltonian matrix in external electric and magnetic fields built from $|\Psi \rangle$ is a $4N_y \times 4N_y$ Hermitian matrix, which can be viewed as a four-block matrix
\begin{equation}
H=\left[
\begin{array}{ c c }
(H_{ij})_{11} & (H_{ij})_{12} \\
(H_{ij})_{21} & (H_{ij})_{22}
\end{array}
\right] _{4N_{y}\times \,4N_{y}}.
\label{eq:EYBZ2011Hamiltonian}
\end{equation}
Each block is a matrix containing $2N_y\times 2N_y$ elements.
The diagonal block matrices are due to the intralayer interactions while the off-diagonal ones are associated with the interlayer interactions.
The upper triangular matrix elements of $H$ are expressed as
\begin{equation}
(H_{ij})_{11}=\left\{
\begin{array}{ l l l }
\gamma _{6}+U_{1A}(m)                                         & \text{if }j=i,     & i=1,3,5,...,2N_{y}-1, \\
U_{1B}(m)                                                              & \text{if }j=i,     & i=2,4,6,...,2N_{y}, \\
2\gamma_{0}\cos [\sqrt{3}bk_{x}/{2}-\Phi _{P1}(m)]  & \text{if }j=i+1, & i=1,3,5,...,2N_{y}-1, \\
\gamma_{0}                                                            & \text{if }j=i+1, & i=2,4,6,...,2N_{y}-2, \\
0                                                                            & \text{otherwise,}
\end{array}
\right.
\label{eq:EYBZ2011H11}
\end{equation}
\begin{equation}
(H_{ij})_{22}=\left\{
\begin{array}{ l l l }
\gamma_{6}+U_{2A}(m)                                          & \text{if }j=i,     & i=1,3,5,...,2N_{y}-1, \\
U_{2B}(m)                                                              & \text{if }j=i,     & i=2,4,6,...,2N_{y}, \\
2\gamma_{0}\cos [\sqrt{3}bk_{x}/{2}-\Phi _{P2}(m)]  & \text{if }j=i+1, & i=1,3,5,...,2N_{y}-1, \\
\gamma_{0}                                                            & \text{if }j=i+3, & i=1,3,5,...,2N_{y}-3, \\
0                                                                            & \text{otherwise,}
\end{array}
\right.
\label{eq:EYBZ2011H22}
\end{equation}
\begin{equation}
(H_{ij})_{12}=\left\{
\begin{array}{ l l l }
\gamma_{1}                                                           & \text{if }j=i,     & i=1,3,5,...,2N_{y}-1, \\
\gamma_{3}                                                           & \text{if }j=i,     & i=2,4,6,...,2N_{y}, \\
2\gamma_{3}\cos [\sqrt{3}bk_{x}/{2}-\Phi _{P3}(m)]  & \text{if }j=i+2, & i=2,4,6,...,2N_{y}-2, \\
2\gamma_{4}\cos [\sqrt{3}bk_{x}/{2}-\Phi _{P1}(m)]  & \text{if }j=i-1,  & i=2,4,6,...,2N_{y}, \\
2\gamma_{4}\cos [\sqrt{3}bk_{x}/{2}-\Phi _{P2}(m)]  & \text{if }j=i+1, & i=1,3,5,...,2N_{y}-1, \\
\gamma_{4}                                                            & \text{if }j=i+3, & i=1,3,5,...,2N_{y}-3, \\
\gamma_{4}                                                            & \text{if }j=i+1, & i=2,4,6,...,2N_{y}-2, \\
0                                                                            & \text{otherwise,}
\end{array}
\right.
\label{eq:EYBZ2011H12}
\end{equation}
where
\begin{subequations} \label{eq:EYBZ2011Uy}
\begin{align}
U_{1A}(m) = & -eF_{y}  (m-1)(3b/2),
\label{eq:EYBZ2011Uy a}\\
U_{1B}(m) = & -eF_{y} [(m-1)(3b/2)+b/2],
\label{eq:EYBZ2011Uy b}\\
U_{2A}(m) = & -eF_{y}  (m-1)(3b/2),
\label{eq:EYBZ2011Uy c}\\
U_{2B}(m) = & -eF_{y} [(m-1)(3b/2)-b/2],
\label{eq:EYBZ2011Uy d}
\end{align}
\end{subequations}
and
\begin{subequations} \label{eq:EYBZ2011PeierlsPhase}
\begin{align}
\Phi_{P1}(m) = &\ \pi \left[ m-\frac{N_{y}+1}{2}\right] \frac{B_{z}}{\phi _{0}} \frac{3\sqrt{3}}{2}b^{2},
\label{eq:EYBZ2011PeierlsPhase a}\\
\Phi_{P2}(m) = &\ \pi \left[ \left( m-\frac{N_{y}+1}{2}\right) - \frac{1}{3}\right] \frac{B_{z}}{\phi _{0}}\frac{3\sqrt{3}}{2}b^{2},
\label{eq:EYBZ2011PeierlsPhase b}\\
\Phi_{P3}(m) = &\ \pi \left[ \left( m-\frac{N_{y}+1}{2}\right) + \frac{1}{3}\right] \frac{B_{z}}{\phi _{0}}\frac{3\sqrt{3}}{2}b^{2},
\label{eq:EYBZ2011PeierlsPhase c}
\end{align}
\end{subequations}
and $m$ ($=1,2,...,N_y$) is the $m$-th zigzag line where the atom sits.
The factor $-(N_y+1)/2$ keeps the band structure symmetric about $k_x=0$ for an arbitrary magnetic flux.
This indicates that the origin of the $x$-axis is set at the center of nanoribbon.
The dangling bonds on the edge sites are assumed to be passivated by hydrogen atoms, which make no contribution to the electronic states near the Fermi level.
The state energy $E^{c,v}(k_x)$ and wave functions are obtained by diagonalizing the Hamiltonian matrix, where the superscript $c$ ($v$) represents unoccupied $\pi^{\ast }$ band (occupied $\pi$ band).

\section{Results and discussions}

\begin{figure}
\begin{center}
  \includegraphics[]{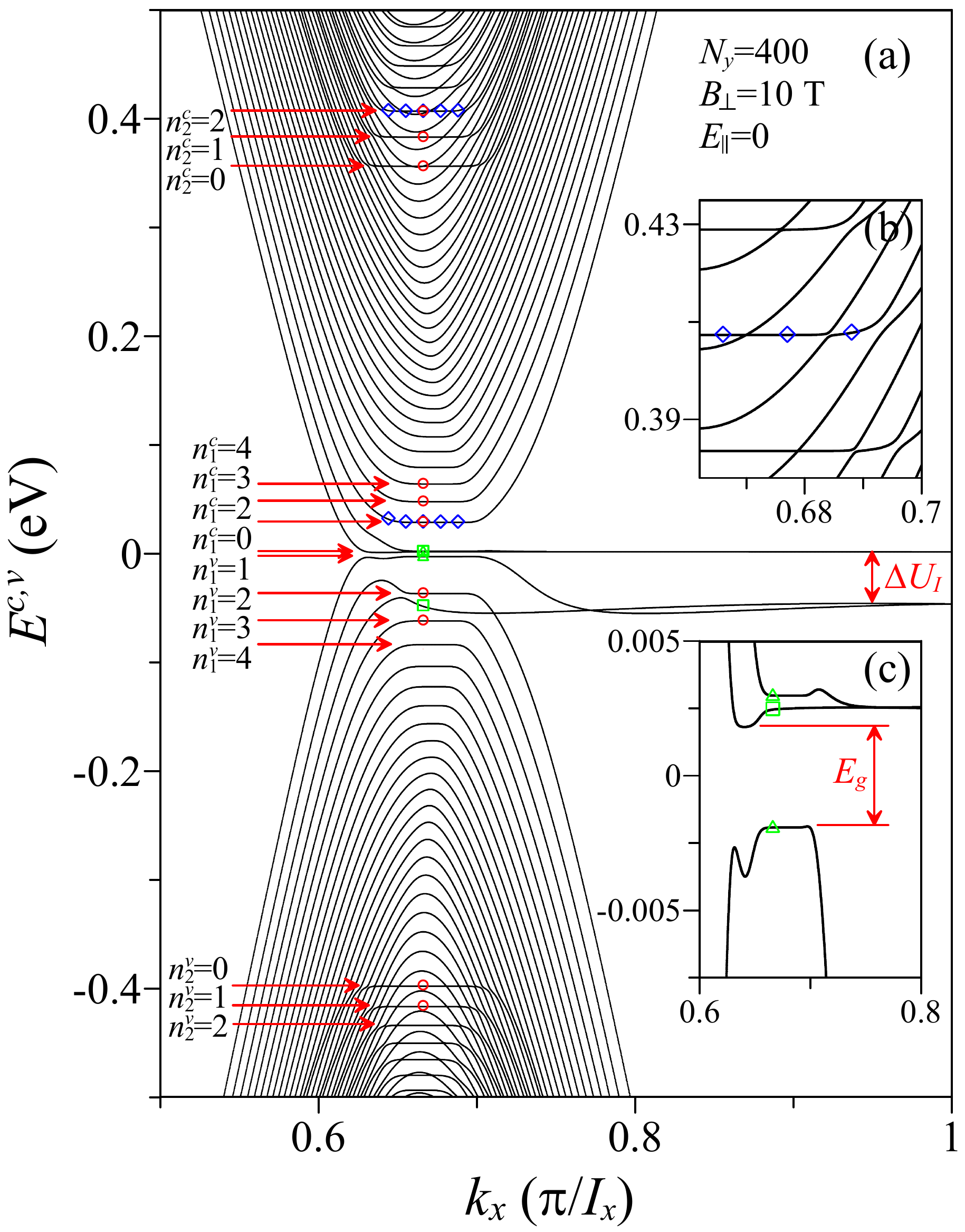}\\
  \caption{
  (a) Band structure of $N_y=400$ bilayer zigzag graphene nanoribbon in magnetic field $B_\perp = 10$ T.
  (b) Band mixing occurs in the overlapping region.
  The $\diamond$ marks from left to right are related to $k_x = k_3$, $k_4$, and $k_5$.
  (c) Energy gap $E_g = 3.7$ meV.
  }
  \label{fig:EYBZ2011BSBilayerZGNRinB}
\end{center}
\end{figure}

The energy spectrum of bilayer zigzag graphene nanoribbon under the influence of magnetic field is depicted in Fig.~\ref{fig:EYBZ2011BSBilayerZGNRinB}.
Only the spectrum in positive $k_x$ region is discussed due to the symmetry of the spectrum about the $k_x=0$ axis.
There are two groups of conduction and valence QLLs and four flat subbands.
Unlike the monolayer graphene nanoribbon, the conduction and valence ones are not symmetric about the Fermi level.
The first group of QLLs occurs from $E_F=0$ and the dispersionless region spreads from $k_x=2\pi /3$.
For QLL with higher state energy $|E^{c,v}|$, the dispersionless region gradually reduces and eventually disappears as the energy is sufficiently large.
These behaviors result from the competition between the confinements of magnetic field and finite width of nanoribbon.
The second group, occurring from $E^{c,v}=0.36$ and $-0.4$ eV, has a similar behavior as the first group.
As the state energy increases, the width of QLL gradually shrinks, and then disappears.
On the other hand, near the Fermi energy, there are four flat subbands, which are twice that revealed in the monolayer zigzag garphene nanoribbon~\cite{J.Phys.Soc.Jpn.80(2011)044602H.C.Chung}.
According to the wave functions (will be discussed latter) at $k_x=2\pi /3$, Landau states and localized states are mixed in the four flat subbands.
It is obvious that interlayer interactions lead to two groups of QLLs occurring from different energies, the symmetry destruction of the unoccupied and occupied QLLs about $E_F=0$, doubling the number of flat subbands, and the splitting of flat subbands with a splitting energy $\Delta U_I=0.048$ eV at $k_x=\pi$.
Furthermore, the band mixing phenomena happen in the overlapping region between two QLL groups [Fig.~\ref{fig:EYBZ2011BSBilayerZGNRinB}(b)], and an energy gap $E_g = 3.7$ meV exists [Fig.~\ref{fig:EYBZ2011BSBilayerZGNRinB}(c)].

\begin{figure}
\begin{center}
  \includegraphics[]{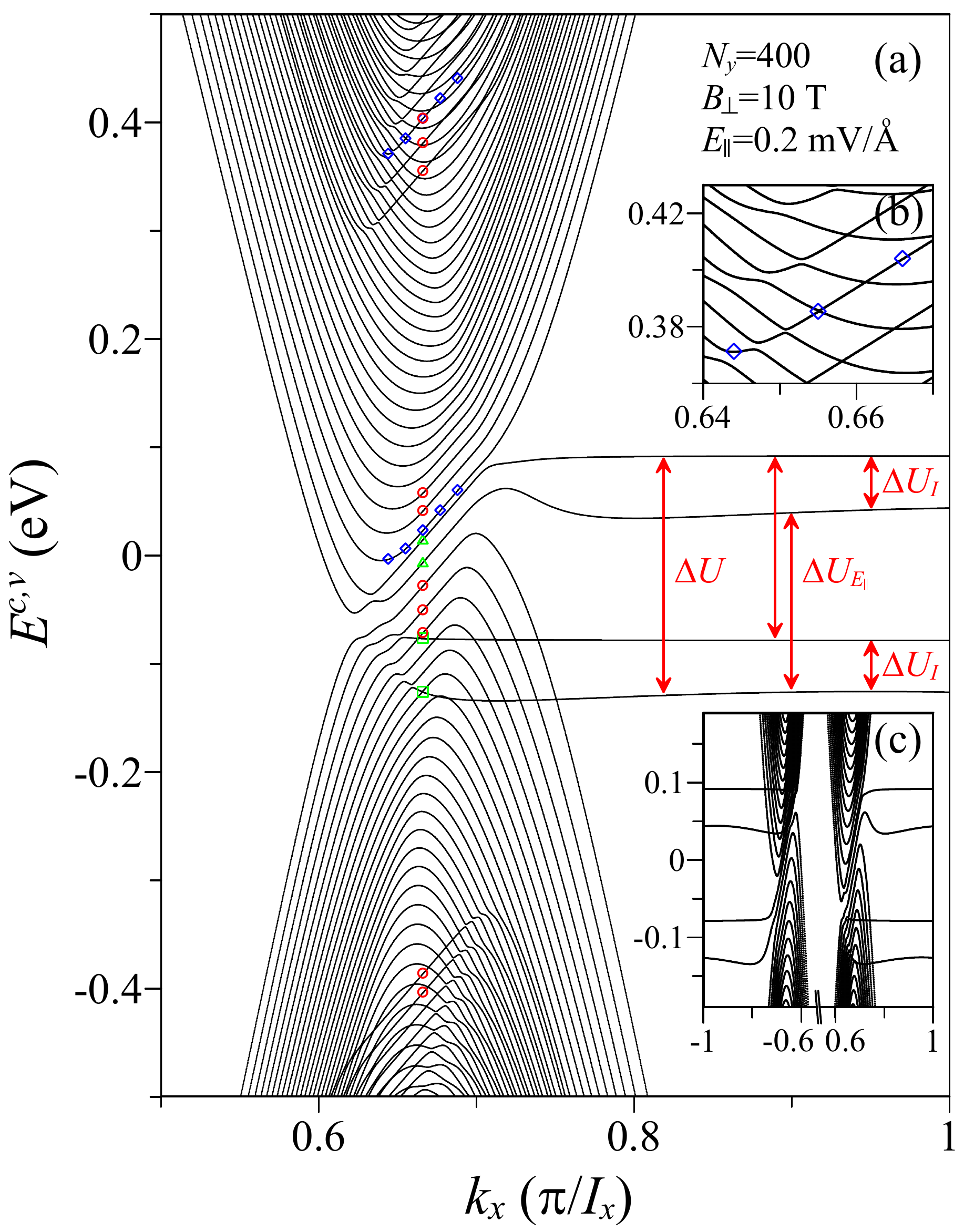}\\
  \caption{
  (a) Band structure of $N_y=400$ bilayer zigzag graphene nanoribbon in magnetic field $B_\perp = 10$ T and electric field $E_\parallel = 0.2$ mV/\AA .
  (b) Band mixing occurs in the overlapping region.
  The $\diamond$ marks from left to right are related to $k_x = k_1$, $k_2$, and $k_3$.
  (c) Low-energy energy spectrum in first Brillouin zone.
  }
  \label{fig:EYBZ2011BSBilayerZGNRinBandE}
\end{center}
\end{figure}

When a transverse electric field is applied, the electrons experience a smoothly changed potential, and the energy spectrum is distorted as shown in Fig.~\ref{fig:EYBZ2011BSBilayerZGNRinBandE}.
The flat dispersionless QLLs are tilted, and the points of QLLs at $k_x=2\pi /3$ are the pivots.
Both groups of QLLs are oblique with the same angle, and the linear shape of each QLL remains.
Although the transverse electric field causes a potential difference for electrons on different zigzag lines, the confinement of magnetic field still holds.
As the subbands tilt, the energy gap shrinks to zero, and a semiconductor-metal transition happens.
The energy spacings of flat subbands are shown in Fig.~\ref{fig:EYBZ2011BSBilayerZGNRinBandE} with $\Delta U= \Delta U_{E_\parallel} + \Delta U_I$, where $\Delta U$ is the total energy spacing and $\Delta U_{E_\parallel} = |e E_\parallel W|$ is the effective energy spacing due to the transverse electric field.
It should be noted that two of the flat subbands tilt and behave like QLLs, and the other two only have an energy shift.
Moreover, the band mixing remains in the overlap region [Fig.~\ref{fig:EYBZ2011BSBilayerZGNRinBandE}(b)], and causes many band edge states, which are corresponding to the peaks in DOS.
For monolayer graphene nanoribbons, the band structures are anti-symmetric at the origin [$(E^{c,v}, k_x)=(0, 0)$], while for bilayer graphene nanoribbons, this anti-symmetry property no longer exists [see Fig.~\ref{fig:EYBZ2011BSBilayerZGNRinBandE}(c)], which indicates more peak appearances in DOS.

Wave functions give more information on the electronic states and a different perspective on the system, such as state mixing, variation of localized and Landau states, as well as distribution of electrons.
They are discussed as follows.
The wave functions of the first and the second groups QLLs at $k_x=2\pi /3$, where the Landau states lie at the ribbon center, are discussed first.
Next is the wave functions of flat subbands at $k_x=2\pi /3$, and the last part focuses on the wave functions of a specific QLL at various wavevectors.

It is convenient to analyze the wave functions by decomposing them as
\begin{eqnarray}
\nonumber
|\Psi \rangle = \sum\limits_{odd}
\bigg[ 1A_o|a_o\rangle +1B_o|b_o\rangle + 2A_o|a_o\rangle +2B_o|b_o\rangle \bigg]\\
\nonumber
                    + \sum\limits_{even}
\bigg[ 1A_e|a_e\rangle +1B_e|b_e\rangle + 2A_e|a_e\rangle +2B_e|b_e\rangle \bigg] , \\
\label{eq:EYBZ2011BilayerWFdecomposed}
\end{eqnarray}
where $1A_{o(e)}$ [$1B_{o(e)}$] is the amplitude coefficients of atoms $1A$ ($1B$) on the odd (even) zigzag line of the first layer while $2A_{o(e)}$ [$2B_{o(e)}$] belongs to the second layer.
Only $1A_o$, $1B_o$, $2A_o$, and $2B_o$ are depicted because of $1A_o=-1A_e$, $1B_o=-1B_e$, $2A_o=-2A_e$, and $2B_o=-2B_e$ when the ribbon is wide enough ($N_y$ is sufficiently large).

\begin{figure}
\begin{center}
  \includegraphics[]{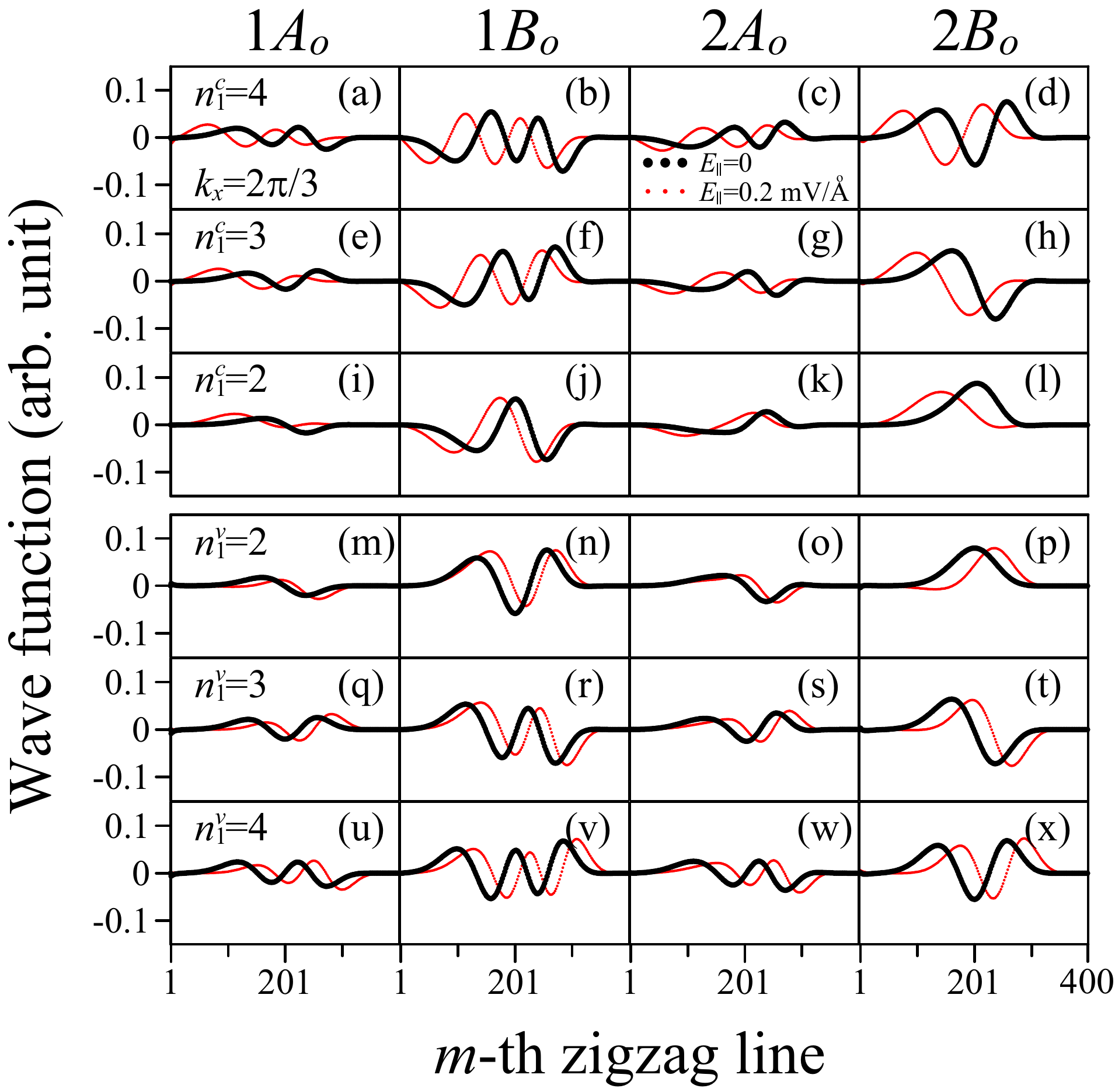}\\
  \caption{
  Wave functions for $n^{c,v}_1=1$--$3$ QLLs at $k_x = 2\pi /3$ ($\circ$ marks in Fig.~\ref{fig:EYBZ2011BSBilayerZGNRinB} and Fig.~\ref{fig:EYBZ2011BSBilayerZGNRinBandE}) in $B_\perp = 10$ T and different strengths of electric fields.
  }
  \label{fig:EYBZ2011WF1stQLLgroup}
\end{center}
\end{figure}

We first focus on the wave functions for the first group of QLLs at $k_x = 2\pi /3$, which will give us a basic view on the Landau wave functions.
Due to the magnetic confinement, the Landau wave functions have regular oscillation patterns (or say nodal structure) at the center of nanoribbon [heavy dots in Fig.~\ref{fig:EYBZ2011WF1stQLLgroup}].
On one hand, for a certain QLL, the nodes (zero points) in sublattice $1A_o$ and $2A_o$ are less than that in sublattice $1B_o$ by one, and the nodes in sublattice $2B_o$ are less than that in sublattice $1B_o$ by two.
For instance, there are 2, 3, 2, and 1 nodes in the sublattices $1A_o$, $1B_o$, $2A_o$, and $2B_o$ of $n^c_1 = 3$ subband, respectively [see Figs.~\ref{fig:EYBZ2011WF1stQLLgroup}(e)--\ref{fig:EYBZ2011WF1stQLLgroup}(h)].
On the other hand, the number of nodes increases regularly as the state energy $|E^{c,v}|$ of QLLs increases.
For example, there are 2, 3, and 4 nodes in the sublattices $1B_o$ of the 2nd, 3rd, and 4th conduction QLLs, respectively [Fig.~\ref{fig:EYBZ2011WF1stQLLgroup}(j), Fig.~\ref{fig:EYBZ2011WF1stQLLgroup}(f), and Fig.~\ref{fig:EYBZ2011WF1stQLLgroup}(b)].
According to the observed regularities and the condition that most electrons are distributed in the $1B_o$ sublattice, the numbers of nodes in sublattices $1B_o$ are used to label the index number of the first group QLLs.

When an electric field is exerted, the Landau wave functions are distorted and shifted.
The wave functions of conduction subbands are shifted to the left side of nanoribbon, which presents an electron-like behavior; while the valence ones are shifted to the right side, which shows a hole-like behavior.
The shift distance is proportional to the state energy of QLLs [light dots in Fig.~\ref{fig:EYBZ2011WF1stQLLgroup}(n), Fig.~\ref{fig:EYBZ2011WF1stQLLgroup}(r), and Fig.~\ref{fig:EYBZ2011WF1stQLLgroup}(v)].
The wave functions are distorted; however, the nodal structure in each sublattice is not changed.

\begin{figure}
\begin{center}
  \includegraphics[]{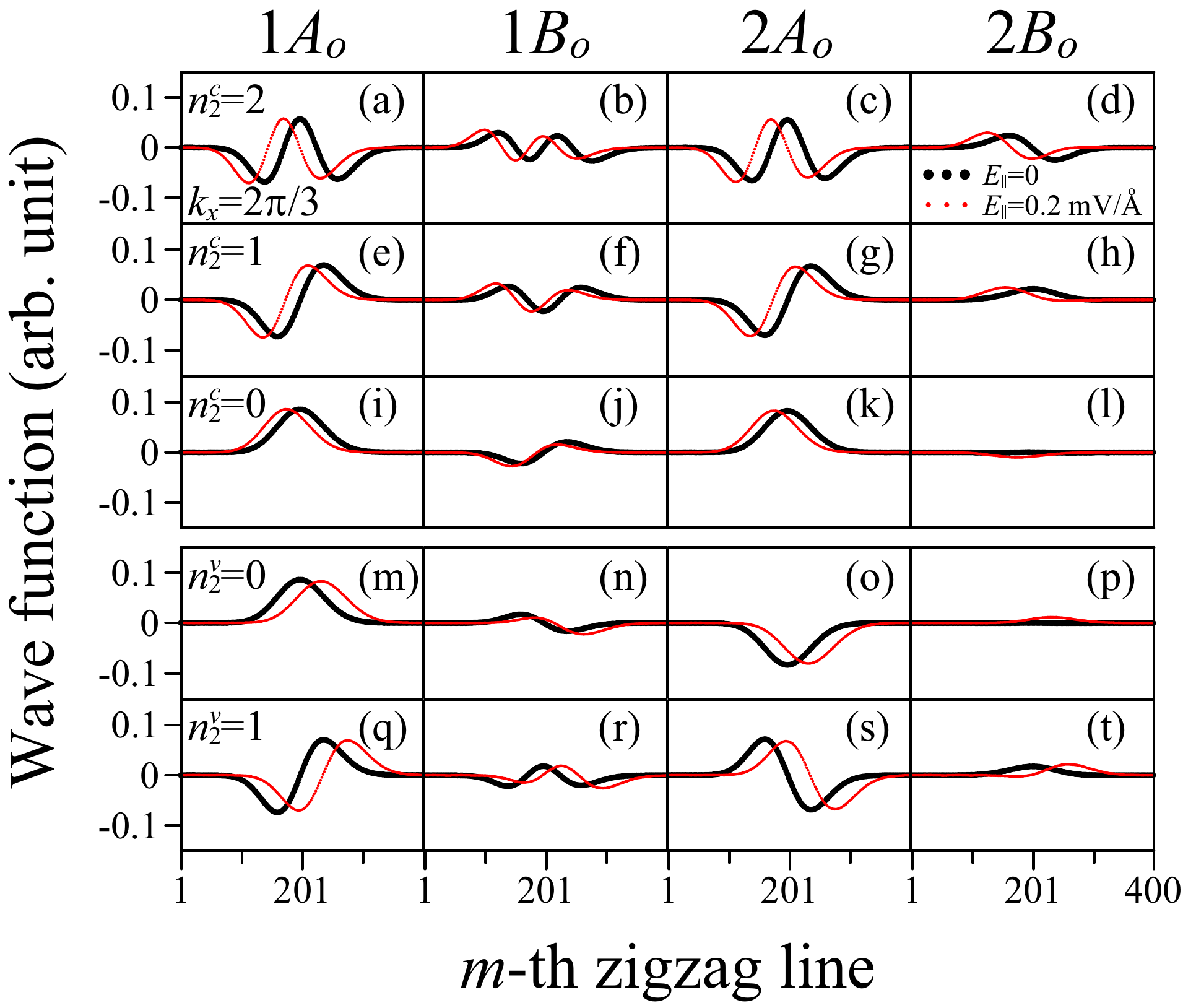}\\
  \caption{
  Wave functions for $n^{c}_2=1$--$3$ and $n^{v}_2=1$, 2 QLLs at $k_x = 2\pi /3$ ($\circ$ marks in Fig.~\ref{fig:EYBZ2011BSBilayerZGNRinB} and Fig.~\ref{fig:EYBZ2011BSBilayerZGNRinBandE}) in $B_\perp = 10$ T and different strengths of electric fields.
  In a certain QLL, the number of nodes in sublattice $1A_o$ ($2A_o$) is less than that in sublattice $1B_o$ by one, and the number of nodes in sublattice $2B_o$ is less than that in sublattice $1B_o$ by two.
  The node numbers increase regularly as the state energy $|E^{c,v}|$ of QLLs increases.
  The $2B_o$ sublattices of $n^{c,v}=0$ subbands are exceptions.
  }
  \label{fig:EYBZ2011WF2ndQLLgroup}
\end{center}
\end{figure}

For the wave functions in the second group of QLLs, although the Landau states remain having regular oscillation patterns, the features are somewhat different.
The wave functions for the second group of QLLs at $k_x = 2\pi /3$ are shown in Fig.~\ref{fig:EYBZ2011WF2ndQLLgroup}.
The regularities in the node number of each sublattice are maintained except for the $2B_o$ sublattices of $n^{c,v}=0$ subbands.
According to the regularities and the condition that most electrons are distributed in the $1A_o$ sublattice, the index of the second group QLLs are labeled by the number of nodes in sublattices $1A_o$.

When an electric field is applied, the Landau wave functions are distorted and shifted in a similar behavior to the first group QLLs.
The wave functions of conduction and valence subbands are shifted to the different side of nanoribbon, and the shift distance is proportional to the state energy of QLLs [light dots in Fig.~\ref{fig:EYBZ2011WF1stQLLgroup}(n), Fig.~\ref{fig:EYBZ2011WF1stQLLgroup}(r), and Fig.~\ref{fig:EYBZ2011WF1stQLLgroup}(v)].
Although the wave functions are distorted, the nodal structure is not affected.

\begin{figure}
\begin{center}
  \includegraphics[]{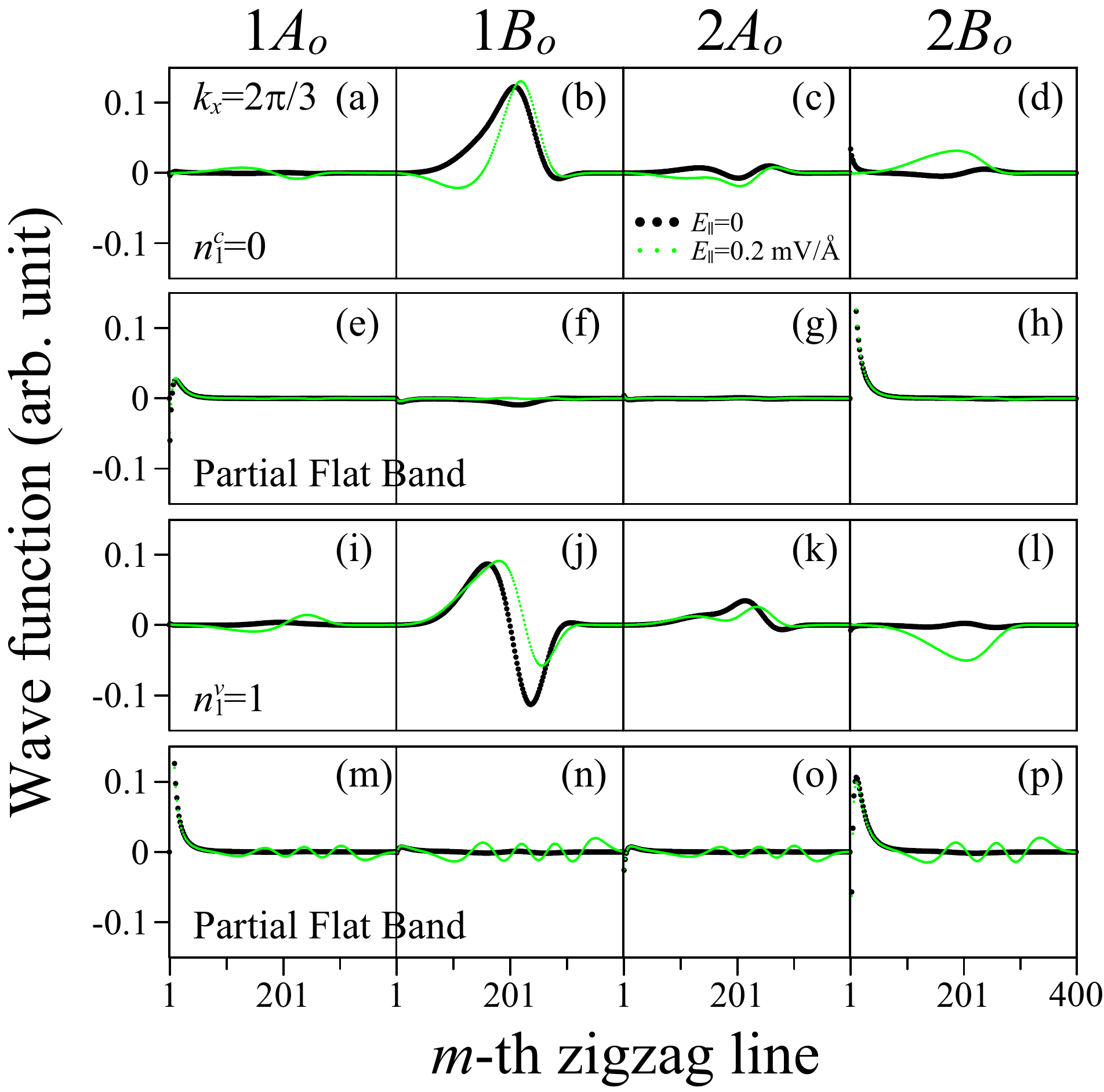}\\
  \caption{
  Wave functions for four flat subbands at $k_x = 2\pi /3$ ($\triangle$ and $\square$ marks in Fig.~\ref{fig:EYBZ2011BSBilayerZGNRinB} and Fig.~\ref{fig:EYBZ2011BSBilayerZGNRinBandE}) in $B_\perp = 10$ T and different strengths of electric fields.
  Each flat subband contains Landau states in the ribbon center and localized states at the ribbon edges.
  The two kinds of states may coexist in a sublattice in $E_\parallel = 0$, e.g., heavy dots in (d).
  After the electric field applies, each subband possesses only one kind of state.
  The oscillations in (m)--(p) are caused by band mixing with the states on $n^v_1=7$ Landau subband.
  }
  \label{fig:EYBZ20116WFofPartialAndLandau}
\end{center}
\end{figure}

Regarding the energy spectra, we know that only two of the flat subbands tilt in the presence of electric field, and the other two don't.
The wave functions give evidence on this conditional change.
In the absence of electric field, the Landau states at the ribbon center coexist with the edge localized states that are from the ribbon edge and identifiable by their exponential decay form.
The coexistence of two kinds of states happens in each sublattice, which is different from the monolayer case.
In monolayer system, the coexistence remains; however, each kind of state occupies a different sublattice~\cite{J.Phys.Soc.Jpn.80(2011)044602H.C.Chung}.
In bilayer system, not only the number of flat subbands is doubled, but also the states have a strong coexistence.
The flat subbands have duality properties on Landau and localized states [shown in Fig.~\ref{fig:EYBZ20116WFofPartialAndLandau}].

When an electric field is exerted, the duality of flat subbands is broken.
Each flat subband contains only Landau states or localized states.
Hence, there is a simple explanation on the two behaviors of the flat subbands.
The subbands with Landau states will tilt as the electric field is applied, and the subbands with localized states will stay flat.
Furthermore, a band mixing phenomenon occurs in the flat subbands.
The states of $n^v_1=7$ Landau subband mix with the states of flat subband, and the nodal structure is very clear.
As you can count, there are 6, 7, 6, and 5 nodes in Figs.~\ref{fig:EYBZ20116WFofPartialAndLandau}(m)--\ref{fig:EYBZ20116WFofPartialAndLandau}(p), respectively.

\begin{figure}
\begin{center}
  \includegraphics[]{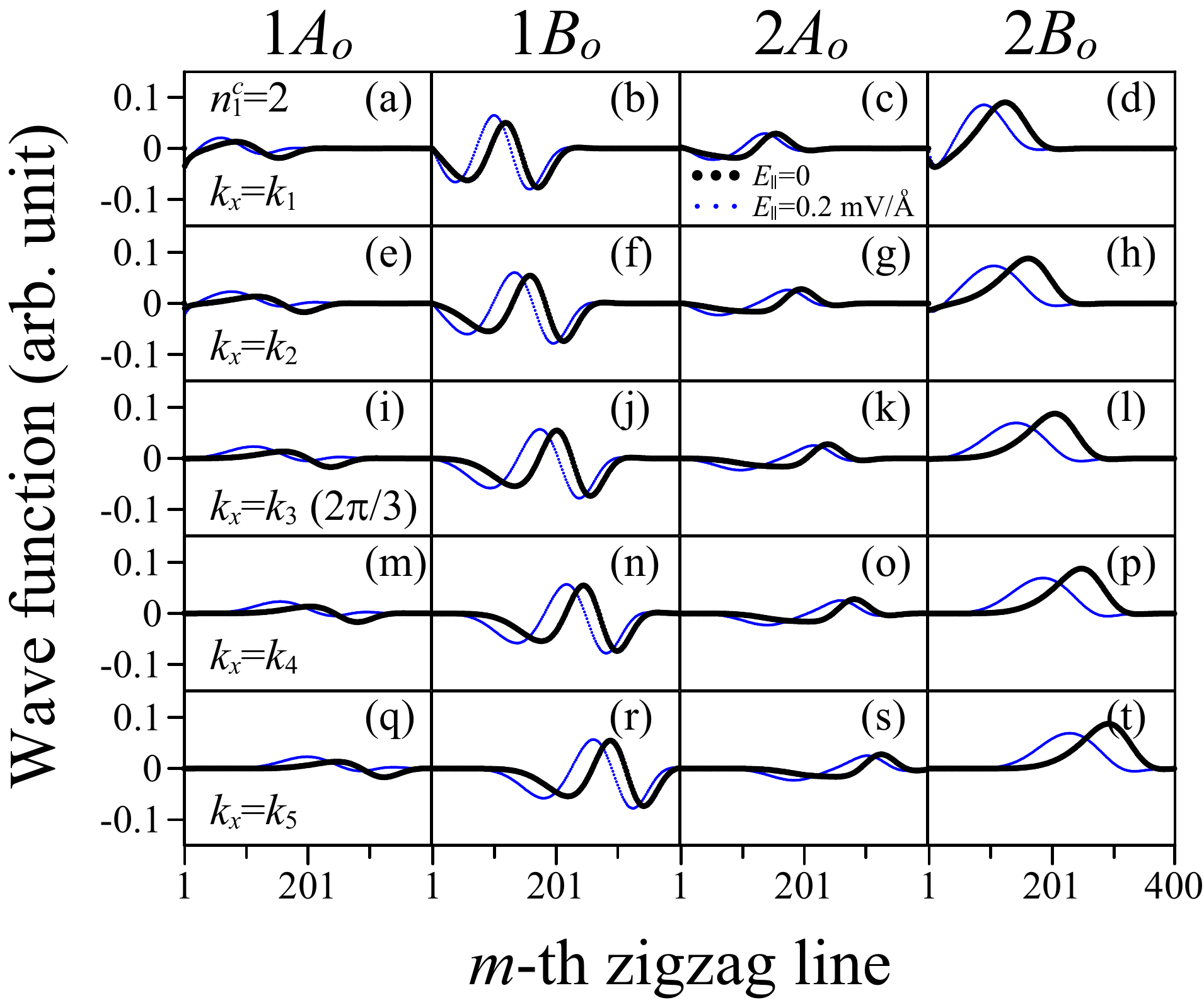}\\
  \caption{
  Wave functions of $n^c_1=2$ subband at different $k_x$ in $B_\perp = 10$ T and different strengths of electric fields.
  The wavevectors $k_x = k_1$--$k_5$ correspond to the $\diamond$ marks on $n^c_1=2$ QLL from left to right in Figs~\ref{fig:EYBZ2011BSBilayerZGNRinB} and~\ref{fig:EYBZ2011BSBilayerZGNRinBandE}.
  }
  \label{fig:EYBZ2011WF1stQLLgroupVSk}
\end{center}
\end{figure}

In the absence of electric field, the location of Landau wave function in the ribbon is associated with the corresponding wavevector.
At $k_x = 2\pi /3$, the Landau wave functions of the first group of QLLs are located at the ribbon center [heavy dots in Figs.~\ref{fig:EYBZ2011WF1stQLLgroupVSk}(i)--(l)].
As the wavevector deviates from $2\pi /3$ ($k_x = k_2$, $k_4$, and $k_5$ in Fig.~\ref{fig:EYBZ2011WF1stQLLgroupVSk}), the wave functions move away from the ribbon center with the shape of wave unchanged.
In addition, when the deviation of $k_x$ is slightly out of the dispersionless region of QLL, the Landau states are distorted as shown in $k_x = k_1$.

As the effect of electric field comes into the system, the Landau wave functions are slightly distorted and pushed to the left side of ribbon with a constant distance [light dots of $k_x = k_3$, $k_4$, and $k_5$ in Fig.~\ref{fig:EYBZ2011WF1stQLLgroupVSk}], which indicates that the positions of cyclotron orbitals are shifted with a constant distance by the electric fields.
For the wave functions whose wavevectors are too close to the edge of QLL (for example, $k_x = k_2$), the cyclotron orbitals are pushed outside the range of tilted QLLs, and the Landau states are broken.

\begin{figure}
\begin{center}
  \includegraphics[]{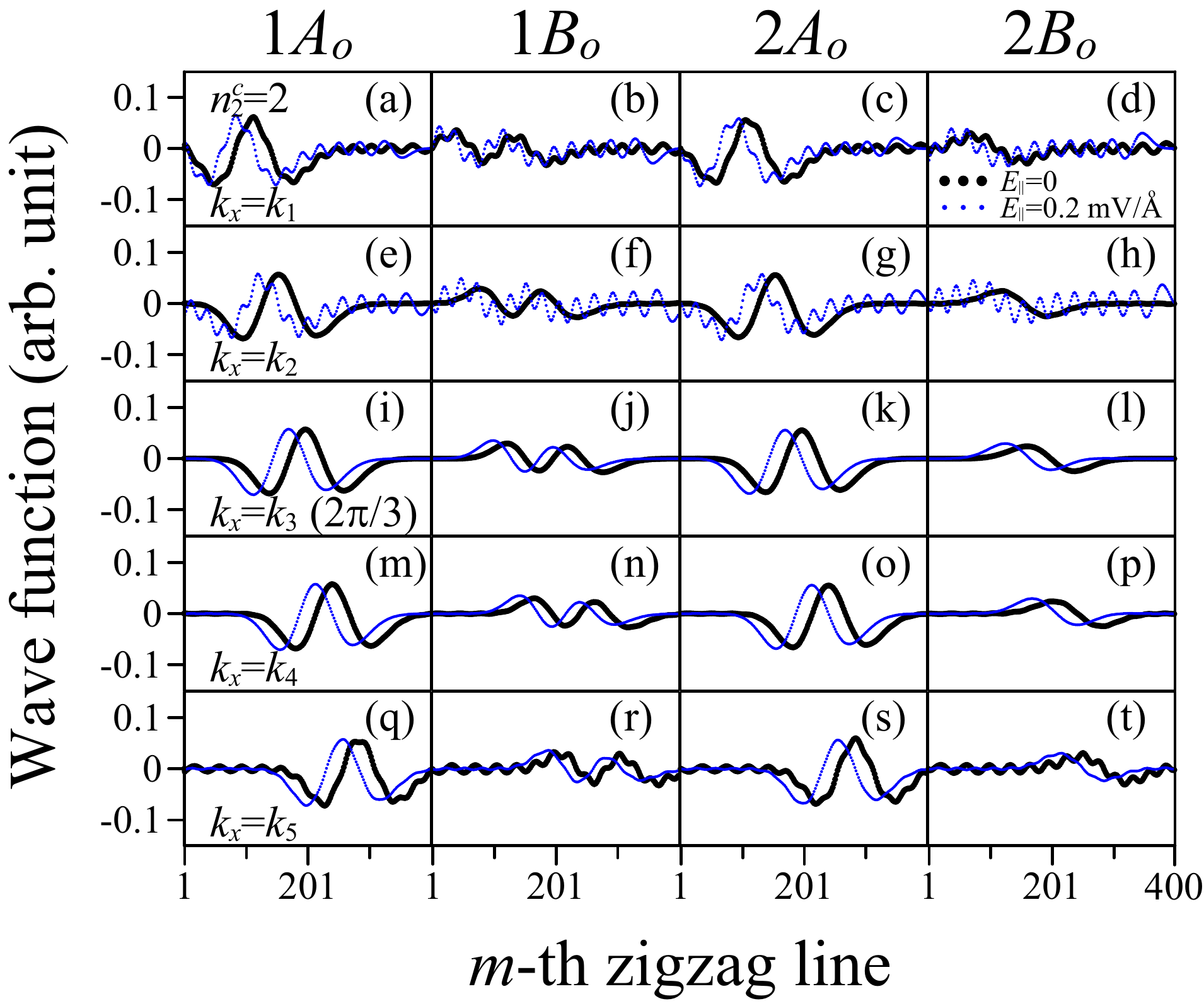}\\
  \caption{
  Wave functions of $n^c_2=2$ subband at different $k_x$ in $B_\perp = 10$ T and different strengths of electric fields.
  The wavevectors $k_x = k_1$--$k_5$ are corresponding to the $\diamond$ remarks on $n^c_2=2$ QLL from left to right in Figs~\ref{fig:EYBZ2011BSBilayerZGNRinB} and~\ref{fig:EYBZ2011BSBilayerZGNRinBandE}.
  The ripples of the wave functions are caused by band mixing with the parabolic subbands.
  }
  \label{fig:EYBZ2011WF2ndQLLgroupVSk}
\end{center}
\end{figure}

For the second group of QLLs in the absence of electric fields, the wave functions at different wavevectors are depicted with heavy dots in Fig.~\ref{fig:EYBZ2011WF2ndQLLgroupVSk}.
As the wavevector deviates from $k_3 = 2\pi /3$, the wave functions move away from the ribbon center.
This behavior is almost the same as the first group, except for band mixing phenomena occurring at the two sides of QLL ($k_x = k_1$ and $k_5$), because many parabolic subbands cross the QLLs as shown in Fig.~\ref{fig:EYBZ2011BSBilayerZGNRinB}(b).

When the electric field is added, the Landau orbitals are shifted a constant distance to the left side of ribbon (see the light dots of $k_x = k_3$ and $k_4$ in Fig.~\ref{fig:EYBZ2011WF2ndQLLgroupVSk}).
In addition, the band mixing phenomena may appear (e.g. $k_x = k_2$) or disappear (e.g. $k_x = k_5$), which is determined from that the tilted QLL crosses the parabolic subbands or not [shown in Fig.~\ref{fig:EYBZ2011BSBilayerZGNRinBandE}(b)].

\begin{figure}
\begin{center}
  \includegraphics[]{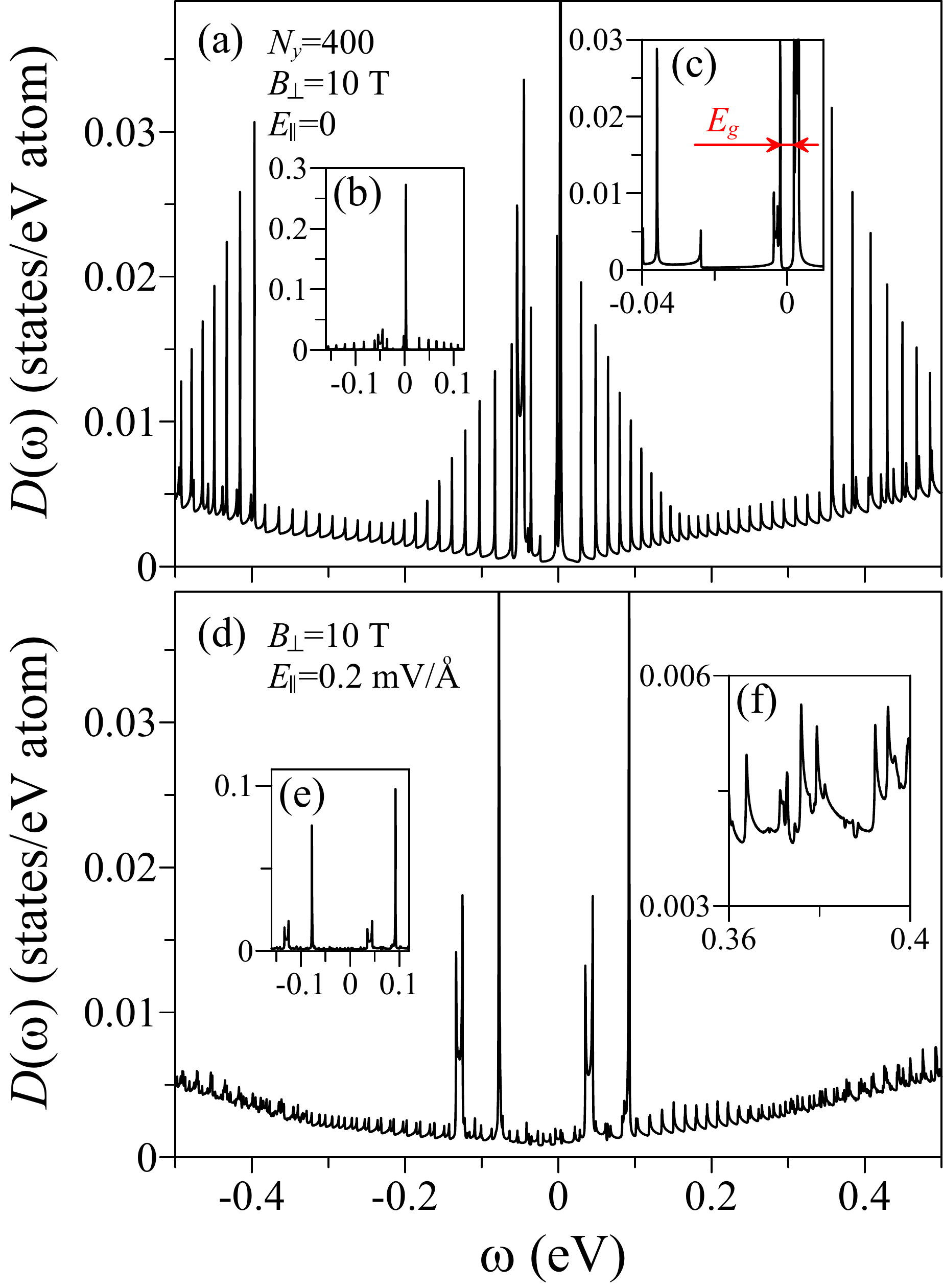}\\
  \caption{
  DOS of Ny = 400 bilayer zigzag graphene nanoribbon in $B_\perp = 10$ T and different strengths of electric fields.
  }
  \label{fig:EYBZ2011DOS}
\end{center}
\end{figure}

DOS directly reflects the main features of energy spectrum, which is given by $D(\omega )= \int dk_x \delta [ E^{c,v}(k_x) -\omega ] $.
In the absence of electric fields, there are two groups of delta-function-like peaks as depicted in Fig.~\ref{fig:EYBZ2011DOS}(a).
The first group starts from frequency $\omega = 0$, and the second group starts from $\omega = 0.36$ and $-0.4$ eV.
The peak position and height are corresponding to the state energy and width of QLLs, respectively.
As the frequency increases, the peak height of the first group decreases rapidly, and the one-side-divergent peaks, which are associated to the parabolic subbands, gradually make their appearances.
At the same time, these peaks start to mix with the second peak group of QLLs.
It is marked in Fig.~\ref{fig:EYBZ2011DOS}(b) that the highest peak are related to the flat subbands near the Fermi level.
The DOS is zero near the Fermi level, which indicates a frequency gap $E_g = 3.7$ meV [Fig.~\ref{fig:EYBZ2011DOS}(c)].

In the presence of electric fields, the two groups of delta-function-like peaks are suppressed [Fig.~\ref{fig:EYBZ2011DOS}(d)].
The delta-function-like peaks become one-side-divergent peaks, since the dispersionless QLLs are tilted.
The six highest peaks are corresponding to the band-edge states of the four flat subbands [shown in Fig.~\ref{fig:EYBZ2011DOS}(e)].
The band mixing phenomenon and the breaking of anti-symmetry in energy spectrum result in many one-side-divergent peaks mixed together [Fig.~\ref{fig:EYBZ2011DOS}(f)].
The features of DOS in this discussion could be verified by the scanning tunneling spectroscopy~\cite{Phys.Rev.B73(2006)085421Y.Niimi, Phys.Rev.Lett.82(1999)1225P.Kim} and the optical absorption spectroscopy~\cite{Science298(2002)2361S.M.Bachilo, Phys.Rev.Lett.90(2003)217401J.Lefebvre}.

\section{Conclusions}

The transverse electric fields cause the distortion of energy spectrum and the semiconductor-metal transition.
Two groups of dispersionless QLLs are tilted, and four flat subbands are split with an effective energy spacing $\Delta U_{E_\parallel} = |e E_\parallel W|$, which directly relates to the strength of electric field and width of nanoribbon.
From the perspective of wave functions, transverse electric fields shift both groups of Landau wave functions with electron-like motion in the conduction QLLs and hole-like motion in the valence QLLs.
The movements of wave functions in any specific QLL are the same, which reflects the uniform strength of the electric field, except for those near the ribbon edges.
For the flat subbands, electric fields completely break the coexistence of Landau and localized states, and this feature is directly reflected on the energy spectra.
Interestingly, more band mixing phenomena, which cause state mixing, occur under the influence of electric fields.
In the DOS, the electric fields suppress all the symmetric peaks of QLLs and the remaining high peaks are related to the flat subbands.
These features in DOS can be verified through surface inspections and optical experiments.

\begin{flushleft}
\textbf{Acknowledgements}
\end{flushleft}

One of us (Hsien-Ching Chung) thanks Ming-Hui Chung and Su-Ming Chen for the financial support.
This work was supported in part by the National Science Council of Taiwan under grant number 98-2112-M-006-013-MY4, and was supported in part by (received funding from) the Headquarters of University Advancement at the National Cheng Kung University, which is sponsored by the Ministry of Education, Taiwan.

\newpage
\bibliography{BilayerEy}

\begin{thebibliography}{10}
\expandafter\ifx\csname url\endcsname\relax
  \def\url#1{\texttt{#1}}\fi
\expandafter\ifx\csname urlprefix\endcsname\relax\def\urlprefix{URL }\fi
\expandafter\ifx\csname href\endcsname\relax
  \def\href#1#2{#2} \def\path#1{#1}\fi

\bibitem{Science306(2004)666K.S.Novoselov}
K.~S. Novoselov, A.~K. Geim, S.~V. Morozov, D.~Jiang, Y.~Zhang, S.~V. Dubonos,
  I.~V. Grigorieva, A.~A. Firsov, \href{<Go to ISI>://000224756700045}{Electric
  field effect in atomically thin carbon films}, Science 306~(5696) (2004)
  666--669.
\newblock \href {http://dx.doi.org/10.1126/science.1102896}
  {\path{doi:10.1126/science.1102896}}.
\newline\urlprefix\url{<Go to ISI>://000224756700045}

\bibitem{Phys.Rev.71(1947)622P.R.Wallace}
P.~R. Wallace, \href{http://link.aps.org/doi/10.1103/PhysRev.71.622}{The band
  theory of graphite}, Physical Review 71~(9) (1947) 622--634.
\newblock \href {http://dx.doi.org/10.1103/PhysRev.71.622}
  {\path{doi:10.1103/PhysRev.71.622}}.
\newline\urlprefix\url{http://link.aps.org/doi/10.1103/PhysRev.71.622}

\bibitem{Nature438(2005)197K.S.Novoselov}
K.~S. Novoselov, A.~K. Geim, S.~V. Morozov, D.~Jiang, M.~I. Katsnelson, I.~V.
  Grigorieva, S.~V. Dubonos, A.~A. Firsov, \href{<Go to
  ISI>://000233133500042}{Two-dimensional gas of massless dirac fermions in
  graphene}, Nature 438~(7065) (2005) 197--200.
\newblock \href {http://dx.doi.org/10.1038/nature04233}
  {\path{doi:10.1038/nature04233}}.
\newline\urlprefix\url{<Go to ISI>://000233133500042}

\bibitem{Nature438(2005)201Y.B.Zhang}
Y.~B. Zhang, Y.~W. Tan, H.~L. Stormer, P.~Kim, \href{<Go to
  ISI>://000233133500043}{Experimental observation of the quantum hall effect
  and berry's phase in graphene}, Nature 438~(7065) (2005) 201--204.
\newblock \href {http://dx.doi.org/10.1038/nature04235}
  {\path{doi:10.1038/nature04235}}.
\newline\urlprefix\url{<Go to ISI>://000233133500043}

\bibitem{Rev.Mod.Phys.80(2008)1337C.W.J.Beenakker}
C.~W.~J. Beenakker, \href{<Go to ISI>://WOS:000262253500007}{Colloquium:
  Andreev reflection and klein tunneling in graphene}, Reviews of Modern
  Physics 80~(4) (2008) 1337--1354.
\newblock \href {http://dx.doi.org/10.1103/RevModPhys.80.1337}
  {\path{doi:10.1103/RevModPhys.80.1337}}.
\newline\urlprefix\url{<Go to ISI>://WOS:000262253500007}

\bibitem{Phys.Rev.Lett.102(2009)026807N.Stander}
N.~Stander, B.~Huard, D.~Goldhaber-Gordon, \href{<Go to
  ISI>://WOS:000262535900059}{Evidence for klein tunneling in graphene p-n
  junctions}, Physical Review Letters 102~(2) (2009) 026807.
\newblock \href {http://dx.doi.org/10.1103/PhysRevLett.102.026807}
  {\path{doi:10.1103/PhysRevLett.102.026807}}.
\newline\urlprefix\url{<Go to ISI>://WOS:000262535900059}

\bibitem{Synth.Met.161(2011)2249Y.Y.Tan}
Y.~Y. Tan, L.~W. Tan, K.~D. G.~I. Jayawardena, J.~V. Anguita, J.~D. Carey,
  S.~R.~P. Silva, \href{<Go to ISI>://WOS:000297878100011}{Field effect in
  chemical vapour deposited graphene incorporating a polymeric gate
  dielectric}, Synthetic Metals 161~(21-22) (2011) 2249.
\newblock \href {http://dx.doi.org/10.1016/j.synthmet.2011.08.029}
  {\path{doi:10.1016/j.synthmet.2011.08.029}}.
\newline\urlprefix\url{<Go to ISI>://WOS:000297878100011}

\bibitem{Science319(2008)1229X.L.Li}
X.~L. Li, X.~R. Wang, L.~Zhang, S.~W. Lee, H.~J. Dai, \href{<Go to
  ISI>://WOS:000253530600039}{Chemically derived, ultrasmooth graphene
  nanoribbon semiconductors}, Science 319~(5867) (2008) 1229--1232.
\newblock \href {http://dx.doi.org/10.1126/science.1150878}
  {\path{doi:10.1126/science.1150878}}.
\newline\urlprefix\url{<Go to ISI>://WOS:000253530600039}

\bibitem{Nature458(2009)872D.V.Kosynkin}
D.~V. Kosynkin, A.~L. Higginbotham, A.~Sinitskii, J.~R. Lomeda, A.~Dimiev,
  B.~K. Price, J.~M. Tour, \href{<Go to
  ISI>://WOS:000265182500039}{Longitudinal unzipping of carbon nanotubes to
  form graphene nanoribbons}, Nature 458~(7240) (2009) 872--876.
\newblock \href {http://dx.doi.org/10.1038/nature07872}
  {\path{doi:10.1038/nature07872}}.
\newline\urlprefix\url{<Go to ISI>://WOS:000265182500039}

\bibitem{Rev.Mod.Phys.81(2009)109A.H.CastroNeto}
A.~H. Castro~Neto, F.~Guinea, N.~M.~R. Peres, K.~S. Novoselov, A.~K. Geim,
  \href{<Go to ISI>://WOS:000264764600004}{The electronic properties of
  graphene}, Reviews of Modern Physics 81~(1) (2009) 109--162.
\newblock \href {http://dx.doi.org/10.1103/RevModPhys.81.109}
  {\path{doi:10.1103/RevModPhys.81.109}}.
\newline\urlprefix\url{<Go to ISI>://WOS:000264764600004}

\bibitem{Phys.Rev.B59(1999)8271K.Wakabayashi}
K.~Wakabayashi, M.~Fujita, H.~Ajiki, M.~Sigrist, \href{<Go to
  ISI>://WOS:000079514300069}{Electronic and magnetic properties of
  nanographite ribbons}, Physical Review B 59~(12) (1999) 8271--8282.
\newline\urlprefix\url{<Go to ISI>://WOS:000079514300069}

\bibitem{J.Phys.Soc.Jpn.80(2011)044602H.C.Chung}
H.-C. Chung, M.-H. Lee, C.-P. Chang, Y.-C. Huang, M.-F. Lin, \href{<Go to
  ISI>://WOS:000289346600023}{Effects of transverse electric fields on
  quasi-landau levels in zigzag graphene nanoribbons}, Journal of the Physical
  Society of Japan 80~(4) (2011) 044602.
\newblock \href {http://dx.doi.org/10.1143/jpsj.80.044602}
  {\path{doi:10.1143/jpsj.80.044602}}.
\newline\urlprefix\url{<Go to ISI>://WOS:000289346600023}

\bibitem{Nanotechnology18(2007)495401Y.C.Huang}
Y.~C. Huang, C.~P. Chang, M.~F. Lin, \href{<Go to
  ISI>://WOS:000252148900006}{Magnetic and quantum confinement effects on
  electronic and optical properties of graphene ribbons}, Nanotechnology
  18~(49) (2007) 495401.
\newblock \href {http://dx.doi.org/10.1088/0957-4484/18/49/495401}
  {\path{doi:10.1088/0957-4484/18/49/495401}}.
\newline\urlprefix\url{<Go to ISI>://WOS:000252148900006}

\bibitem{PhysicaE42(2010)711H.C.Chung}
H.~C. Chung, Y.~C. Huang, M.~H. Lee, C.~C. Chang, M.~F. Lin, \href{<Go to
  ISI>://WOS:000276541200012}{Quasi-landau levels in bilayer zigzag graphene
  nanoribbons}, Physica E-Low-Dimensional Systems and Nanostructures 42~(4)
  (2010) 711--714.
\newblock \href {http://dx.doi.org/10.1016/j.physe.2009.11.090}
  {\path{doi:10.1016/j.physe.2009.11.090}}.
\newline\urlprefix\url{<Go to ISI>://WOS:000276541200012}

\bibitem{Opt.Express19(2011)23350H.C.Chung}
H.~C. Chung, M.~H. Lee, C.~P. Chang, M.~F. Lin, \href{<Go to
  ISI>://WOS:000296904700115}{Exploration of edge-dependent optical selection
  rules for graphene nanoribbons}, Optics Express 19~(23) (2011) 23350.
\newblock \href {http://dx.doi.org/10.1364/OE.19.023350}
  {\path{doi:10.1364/OE.19.023350}}.
\newline\urlprefix\url{<Go to ISI>://WOS:000296904700115}

\bibitem{J.Phys.Soc.Jpn.69(2000)3529M.F.Lin}
M.~F. Lin, F.~L. Shyu, \href{<Go to ISI>://WOS:000165569400015}{Optical
  properties of nanographite ribbons}, Journal of the Physical Society of Japan
  69~(11) (2000) 3529--3532.
\newline\urlprefix\url{<Go to ISI>://WOS:000165569400015}

\bibitem{J.Phys.Soc.Jpn.70(2001)2513M.F.Lin}
M.~F. Lin, M.~Y. Chen, F.~L. Shyu, \href{<Go to
  ISI>://WOS:000171216700006}{Electronic collective excitations in ab-stacked
  nanographite ribbons}, Journal of the Physical Society of Japan 70~(9) (2001)
  2513--2516.
\newblock \href {http://dx.doi.org/10.1143/JPSJ.70.2513}
  {\path{doi:10.1143/JPSJ.70.2513}}.
\newline\urlprefix\url{<Go to ISI>://WOS:000171216700006}

\bibitem{Phys.Rev.B76(2007)205433A.Cresti}
A.~Cresti, G.~Grosso, G.~P. Parravicini, \href{<Go to
  ISI>://WOS:000251326900106}{Numerical study of electronic transport in gated
  graphene ribbons}, Physical Review B 76~(20) (2007) 205433.
\newblock \href {http://dx.doi.org/10.1103/PhysRevB.76.205433}
  {\path{doi:10.1103/PhysRevB.76.205433}}.
\newline\urlprefix\url{<Go to ISI>://WOS:000251326900106}

\bibitem{Synth.Met.171(2013)7T.S.Li}
T.~S. Li, M.~F. Lin, C.~Y. Lin, S.~C. Chang, S.~P. Yang, \href{<Go to
  ISI>://WOS:000319848600002}{Electronic properties of curved graphene
  nanoribbons}, Synthetic Metals 171 (2013) 7.
\newblock \href {http://dx.doi.org/10.1016/j.synthmet.2013.02.022}
  {\path{doi:10.1016/j.synthmet.2013.02.022}}.
\newline\urlprefix\url{<Go to ISI>://WOS:000319848600002}

\bibitem{Carbon42(2004)2975C.P.Chang}
C.~P. Chang, C.~L. Lu, F.~L. Shyu, R.~B. Chen, Y.~K. Fang, M.~F. Lin, \href{<Go
  to ISI>://WOS:000224331300021}{Magneto electronic properties of a graphite
  sheet}, Carbon 42~(14) (2004) 2975--2980.
\newblock \href {http://dx.doi.org/10.1016/j.carbon.2004.07.011}
  {\path{doi:10.1016/j.carbon.2004.07.011}}.
\newline\urlprefix\url{<Go to ISI>://WOS:000224331300021}

\bibitem{Phys.Rev.B43(1991)4579J.C.Charlier}
J.~C. Charlier, X.~Gonze, J.~P. Michenaud, \href{<Go to
  ISI>://WOS:A1991EY62300003}{First-principles study of the electronic
  properties of graphite}, Physical Review B 43~(6) (1991) 4579--4589.
\newblock \href {http://dx.doi.org/10.1103/PhysRevB.43.4579}
  {\path{doi:10.1103/PhysRevB.43.4579}}.
\newline\urlprefix\url{<Go to ISI>://WOS:A1991EY62300003}

\bibitem{PhysicalPropertiesOfCarbonNanotubes(1998)p98R.Saito}
R.~Saito, G.~Dresselhaus, M.~S. Dresselhaus, Physical Properties of Carbon
  Nanotubes, World Scientific Publishing Company, 1998, book section~6, p.~98.

\bibitem{Phys.Rev.B73(2006)085421Y.Niimi}
Y.~Niimi, T.~Matsui, H.~Kambara, K.~Tagami, M.~Tsukada, H.~Fukuyama, \href{<Go
  to ISI>://WOS:000235669100090}{Scanning tunneling microscopy and spectroscopy
  of the electronic local density of states of graphite surfaces near
  monoatomic step edges}, Physical Review B 73~(8) (2006) 085421.
\newblock \href {http://dx.doi.org/10.1103/PhysRevB.73.085421}
  {\path{doi:10.1103/PhysRevB.73.085421}}.
\newline\urlprefix\url{<Go to ISI>://WOS:000235669100090}

\bibitem{Phys.Rev.Lett.82(1999)1225P.Kim}
P.~Kim, T.~W. Odom, J.~L. Huang, C.~M. Lieber, \href{<Go to
  ISI>://WOS:000078458500037}{Electronic density of states of atomically
  resolved single-walled carbon nanotubes: Van hove singularities and end
  states}, Physical Review Letters 82~(6) (1999) 1225--1228.
\newline\urlprefix\url{<Go to ISI>://WOS:000078458500037}

\bibitem{Science298(2002)2361S.M.Bachilo}
S.~M. Bachilo, M.~S. Strano, C.~Kittrell, R.~H. Hauge, R.~E. Smalley, R.~B.
  Weisman, \href{<Go to ISI>://WOS:000179915900048}{Structure-assigned optical
  spectra of single-walled carbon nanotubes}, Science 298~(5602) (2002)
  2361--2366.
\newblock \href {http://dx.doi.org/10.1126/science.1078727}
  {\path{doi:10.1126/science.1078727}}.
\newline\urlprefix\url{<Go to ISI>://WOS:000179915900048}

\bibitem{Phys.Rev.Lett.90(2003)217401J.Lefebvre}
J.~Lefebvre, Y.~Homma, P.~Finnie, \href{<Go to
  ISI>://WOS:000183223400048}{Bright band gap photoluminescence from
  unprocessed single-walled carbon nanotubes}, Physical Review Letters 90~(21)
  (2003) 217401.
\newblock \href {http://dx.doi.org/10.1103/PhysRevLett.90.217401}
  {\path{doi:10.1103/PhysRevLett.90.217401}}.
\newline\urlprefix\url{<Go to ISI>://WOS:000183223400048}

\end{thebibliography}
\bibliographystyle{elsarticle-num}

\newpage


\newpage

\newpage

\newpage

\newpage

\newpage

\newpage

\newpage

\newpage

\end{document}